\documentclass[10pt]{iopart}

\usepackage{multicol}
\usepackage{graphicx}
\usepackage[font=small,labelfont=bf]{caption}
\begin{document}

\title[Publications of the Astronomical Society of the Pacific - Dissertation Summary]{Study of fullerene-based molecular nanostructures in planetary nebulae}

\author{J. J. D\'iaz-Luis}

\address{Observatorio Astron\'omico Nacional (IGN), Alfonso XII, 3 y 5, 28014 Madrid, Spain; jjairo@oan.es \\
Thesis work conducted at Instituto de Astrof\'isica de Canarias (IAC) and Departamento de Astrof\'isica, Universidad de La Laguna (ULL), Tenerife, Spain \\
Ph.D. Thesis directed by D. A. Garc\'ia-Hern\'andez and A. Manchado; Ph.D. Degree awarded 2017 September 25
}
%
%
%
%
%

\begin{multicols}{2}
The main goal of this thesis is to unveil some questions related to the formation of complex fullerene-based molecules in space, with the aim of resolving some key problems in astrophysics. The unexpected detections of fullerenes and graphene (possible C$_{24}$) in the H-rich circumstellar environments of evolved stars indicate that these complex molecules are not so rare and bring the idea that other forms of carbon such as hydrogenated fullerenes (fulleranes), buckyonions, and carbon nanotubes may be widespread in the Universe, being closely involved in many aspects of circumstellar/interstellar chemistry and physics. We explore this new and fertile field of research by focusing our study on some Galactic planetary nebulae (PNe) that contain fullerenes. In order to do this, we make use of laboratory spectra of several fullerene-related compounds and compare them with astronomical data. This work is a first step to open up a new field of interdisciplinary research, crossing the boundaries between astronomers, chemists, and physicists, and understand the significant presence of fullerene structures in circumstellar/interstellar environments.

First, we give a complete search of diffuse bands towards three fullerene-rich PNe (Tc 1, M 1-20, and IC 418), and a detailed analysis of the radial velocities of some diffuse interstellar bands (DIBs). In particular, some DIBs are unusually intense towards these objects; for example, we find an unsually strong 4428 \AA\ absorption feature as a common characteristic of fullerene-rich PNe. Interestingly, we report the first possible detection of two diffuse circumstellar bands (DCBs) at 4428 (see Fig. 1) and 5780 \AA\ in the fullerene-rich circumstellar environment around PN Tc 1 (see D\'iaz-Luis et al. 2015 for more details).

\begin{center}
\includegraphics[scale=0.22]{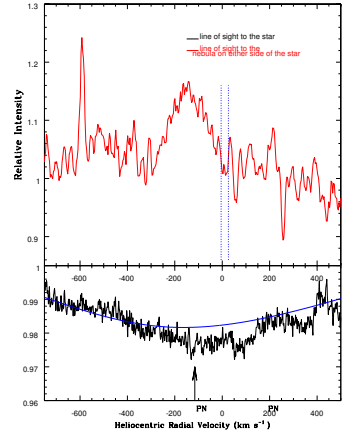}
\captionof{figure}{Broad 4428 $\AA$ band towards Tc 1 central star (black) and average of two sight lines to the nebular position (red). Note the coincidence in velocity (black arrow) of the broad 4428 circumstellar absorption and the corresponding nebular emission.}
\end{center}

Second, we present VLT/ISAAC spectra in the 2.9-4.1 $\mu$m spectral region for the two fullerene-containing PNe Tc 1 and M 1-20. We report the non-detection of the most intense infrared bands of fullerene-related molecules such as fulleranes (like C$_{60}$H$_{36}$ and C$_{60}$H$_{18}$), around $\sim$3.4-3.6 $\mu$m in both PNe. In addition to these non-detections, fulleranes were also tentatively detected in the proto-PN IRAS 01005+7910 (see Zhang \& Kwok 2013), suggesting that fulleranes may be potentially formed in the short transition phase between AGB stars and PNe, and then quickly destroyed by the UV radiation field from the central star (see D\'iaz-Luis et al. 2016 for more details). 
\break \break
Finally, we present narrow-band mid-IR GTC/\-Ca\-na\-ri\-Cam images of the extended fullerene-rich PN IC 418. We study the relative spatial distribution of C$_{60}$- and PAH-like (see Fig. 2) species as well as the 9-13 $\mu$m carrier, with the intention of getting some observational constraints to the formation process of fullerenes in these H-rich environments. Other fullerene-based species (e.g., fulleranes) may contribute to the observed 17.4 $\mu$m emission obtained when subtracting the dust continuum emission (see D\'iaz-Luis et al. 2018 for more details).  

\begin{center}
\includegraphics[scale=0.16]{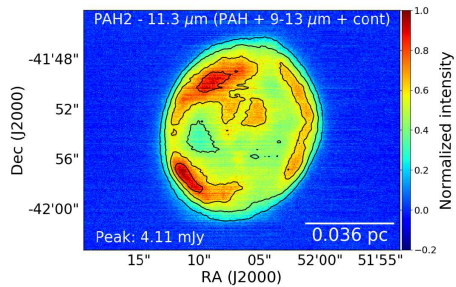}
\captionof{figure}{Contour map of the flux-calibrated mid-IR GTC/\-Ca\-na\-ri\-Cam image of the C$_{60}$-PN IC 418 in the PAH2 (PAH-like $+$ 9-13 $\mu$m $+$ continuum at 11.3 $\mu$m) filter. North is up, east is left. The bar in the lower right corner illustrates 0.036 pc at the estimated distance to IC 418 (1.26 kpc). Contours range from 0.2 to 0.8 with 3 steps of 0.2 each.}
\end{center}

\subsection*{Acknowledgments}
JJDL acknowledges support provided by the Spanish MINECO, grant AYA2016$-$78994$-$P.

\end{multicols}
\end{document}